\documentclass[doublecol,figures]{epl2}
\usepackage{amsmath}
\usepackage{amsfonts}
\usepackage[T1]{fontenc}
\usepackage[english]{babel}
\usepackage{graphicx}
\usepackage{amssymb}
\usepackage{enumitem}
\usepackage{float}

\title{Charged lepton beams as a source of effective neutrinos}

\author{I. Alikhanov\thanks{E-mail: \email{ialspbu@gmail.com}}}
\shortauthor{I. Alikhanov}

\institute{ Institute for Nuclear Research of the Russian Academy of Sciences, 117312 Moscow, Russia\\ 
Institute of Applied Mathematics and Automation KBSC RAS, 360000 Nalchik, Russia
}

\abstract{
Neutrinos are likely the most poorly understood basic constituents of the Standard Model. In order to investigate precisely their interactions one should be able to create high intensity and well-collimated neutrino beams with known flavor compositions. This is a challenging problem for neutrino experiments. We propose a method of studying neutrino interactions based on the fact that a charged lepton is able to manifest itself effectively, with a certain probability, as a neutrino.  The effective neutrino method  may provide an additional tool for probing neutrino-induced reactions at $e^+e^-$ and $ep$ colliders as well as at other facilities that use charged lepton beams. We derive the distributions of the effective neutrinos in the charged leptons and give examples of application of the method to electron--positron collisions.}

\begin{document}

\maketitle

\section{Introduction}
Neutrinos have fundamental implications for various fields of physics: particle physics, astrophysics and cosmology~\cite{King:2014nza,Bilenky:2014ema}. They are the only known particles that have led to an extension  of the Standard Model due to the violation of the law of conservation of the family lepton numbers~\cite{Fukuda:1998ah,Ahmad:2002jz,Eguchi:2002dm} and still display properties requiring explanations. In order to investigate precisely neutrino interactions one should be able to create high intensity and well-collimated neutrino beams with known flavor compositions. This is a challenging problem for neutrino experiments. The currently exploited or discussed techniques are based on the idea of generation of high energy neutrino fluxes through in-flight decays of mesons and boosted beta-radioactive ions \cite{Zucchelli:2002sa,Volpe:2003fi}. 

On the other hand, there are well-established examples of probing dynamical properties of particles that are unavailable in free states at ordinary accelerator experiments. First of all, this is the quark--parton model (QPM)~\cite{Bjorken:1969ja} according to which a hadron can manifest itself as a quark or gluon (parton) carrying the given fraction of the parent hadron momentum. The QPM represents the cross sections for lepton--hadron or hadron--hadron collisions as sums of lepton--quark, quark--quark, gluon--quark or gluon--gluon subprocess cross sections. The observables become thus dependent on the peculiarities of the parton interactions. Another example is the possibility of studying two photon processes at electron--positron colliders~\cite{Brodsky:1971ud,Walsh:1973mz,Budnev:1974de}. Though there is no real photon in the initial state, the electrically charged electron (positron) acts as an effective photon beam according to the Weizs\"acker--Williams approximation. The massive electroweak bosons, $W$ and $Z$, can be treated as partons in the description of various reactions as well~\cite{Kane:1984bb,Dawson:1984gx}. 

We propose to extend the concept of the effective (equivalent) particle also to neutrinos. We show that the distributions of the effective neutrinos in the charged leptons can be derived within a simple approach. The effective neutrino approximation (ENA) may provide a framework for probing neutrino-induced reactions observation of which in laboratory conditions is considered by many unrealistic or at least very difficult. For example, relying on the ENA one can investigate $\nu_e\bar\nu_e$ annihilation into different final states at $e^+e^-$ colliders in the same way as the quark--antiquark annihilation was effectively analyzed through Drell--Yan processes~\cite{Drell:1970wh}.

\section{Neutrino distributions in charged leptons\label{sec1}}
Before proceeding to the discussion of the distributions of the effective neutrinos in the charged leptons, it is useful to begin with the Weizs\"acker--Williams equivalent photon approximation (EPA).  The probability density of finding a photon inside an electron with fraction $x$ of the parent \newline 

\begin{figure}[H]
\centerline{
\includegraphics[width=0.49\textwidth]{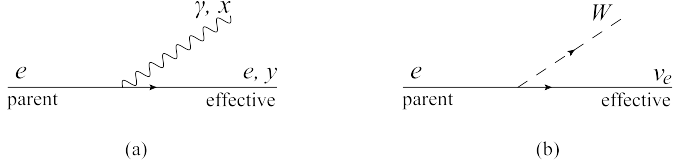}}
\caption{The effective electron in the parent electron (a). The effective electron neutrino in the parent electron (b).}
\label{fig1p}
\end{figure}

\noindent electron momentum can be written as~\cite{Frixione:1993yw}

\begin{equation}
f_{\gamma/e}(x,Q^2)=\frac{\alpha}{2\pi}\left[\frac{1+(1-x)^2}{x}\ln\left(\frac{Q^2_{\text{max}}}{Q^2_{\text{min}}}\right)+\mathcal{O}(1)\right],\label{eq:gen1}
\end{equation}
where $\alpha=e^2/(4\pi)$ is the fine structure constant, $Q^2_{\text{min}}$ and $Q^2_{\text{max}}$ are the minimum and maximum of the magnitude of the four-momentum transfer in the given process. The non-logarithmic term is, in general, also a function of $x$ and $Q^2$. 

It is obvious that in the limit of vanishing electron mass  the outgoing electron, after the emission of the photon, carries the fraction $y=1-x$ of the initial momentum (see Fig.~\ref{fig1p}). This means that~\eqref{eq:gen1} represents simultaneously the probability density of finding an electron with momentum fraction $y$. Therefore the distribution of the effective (equivalent) electrons in the electron is given by

\begin{equation}
f_{e/e}(y,Q^2)=\frac{\alpha}{2\pi}\left[\frac{1+y^2}{1-y}\ln\left(\frac{Q^2_{\text{max}}}{Q^2_{\text{min}}}\right)+\mathcal{O}(1)\right],\label{eq:gen2}
\end{equation}
where  $0\leq y\leq 1$. A detailed discussion of these functions can be found in~\cite{Chen:1975sh}.

In analogy with the EPA, the distribution of the effective $W^-$ bosons in the electron gives simultaneously the distribution of the effective electron neutrinos since the emission of $W^-$ is always accompanied by $\nu_e$ as shown in Fig.~\ref{fig1p}. Hence, relying on the results of~\cite{Alikhanov:2018kpa}, we readily find the distribution of the effective electron neutrinos in the electron to be

\begin{align}
f^T_{\nu/e}(y,Q^2)=\frac{\alpha}{2\pi}\frac{1}{4\sin^2{\theta_W}}\left[\frac{1+\left(y+m_W^2/s\right)^2}{1-y-m_W^2/s}\right.\hspace{1.5cm}\nonumber\\\left.\times\ln\left(\frac{Q^2_{\text{max}}+m_W^2}{Q^2_{\text{min}}+m_W^2}\right)-\frac{1}{s}\left(Q^2_{\text{max}}-Q^2_{\text{min}}\right)\right],
\label{funw}
\end{align}
where the $T$ superscript is to indicate that the accompanied boson is transversely polarized, $\theta_W$ is the Weinberg angle, $s$ is the total center-of-mass (cms) energy of the reaction squared. The non-logarithmic term in the square brackets does not exceed unity because $Q^2_{\text{max}}< s$. Note that $0\leq y\leq 1-m_W^2/s$ as the emitted boson is now massive~\cite{Alikhanov:2018kpa} and the validity of the condition $\sqrt{s}>m_W$ is simultaneously assumed. Also note that, in the high energy limit, \eqref{funw} transforms into the distribution of the effective $W^-$ bosons in the electron with the replacement $y\rightleftarrows (1-y)$, being consistent with the spectrum from~\cite{Kane:1984bb, Dawson:1984gx}.  This is similar to the property that the Weizs\"acker--Williams equivalent photon spectrum of the electron can be obtained from~\eqref{eq:gen2} by simply exchanging the electron and photon momenta~\cite{Chen:1975sh}. One can see that at $m_W\rightarrow 0$ and $1/(4\sin^2{\theta_W})\rightarrow 1$, \eqref{funw} reduces to~\eqref{eq:gen2}.

 We can generalize~\eqref{funw} to the case of the off-shell $W$ bosons of arbitrary mass $q^2\geq0$ as follows:

\begin{align}
\frac{\mathrm{d}f^T_{\nu/e}}{\mathrm{d}q^2}=\frac{\alpha}{2\pi}\frac{1}{4\sin^2{\theta_W}}\left[\frac{1+\left(y+q^2/s\right)^2}{1-y-q^2/s}\right.\hspace{2.3cm}\nonumber\\\left.\times\ln\left(\frac{Q^2_{\text{max}}+q^2}{Q^2_{\text{min}}+q^2}\right)-\frac{1}{s}\left(Q^2_{\text{max}}-Q^2_{\text{min}}\right)\right]\rho(q^2).\nonumber\\
\label{funw1}
\end{align}
Here we have made use of the Breit--Wigner density function corresponding to the $W$ propagator: 
\begin{equation}
\rho(q^2)=\frac{1}{\pi}\frac{\sqrt{q^2}\,\Gamma(q^2)}{(q^2-m_W^2)^2+q^2\Gamma^2(q^2)}
\label{bw_fun}
\end{equation}
with the $W$ width being

\begin{equation}
\Gamma(q^2)=\frac{\alpha}{\sin^2{\theta_W}}\sqrt{q^2}.
\label{bw_fun2}
\end{equation}
Again $0\leq y\leq 1-q^2/s$ and $s>q^2$. In the on-shell limit,~\eqref{bw_fun} gives $\rho(q^2)=\delta(q^2-m_W^2)$, so that~\eqref{funw1} turns into~\eqref{funw}, as it should be.

In the same way as above, adopting the distribution of the effective longitudinally polarized bosons in the electron from~\cite{Alikhanov:2018kpa}, we derive the related effective neutrino distribution:
\begin{equation}
\frac{\mathrm{d}f^L_{\nu/e}}{\mathrm{d}q^2}=\frac{\alpha}{4\pi\sin^2{\theta_W}}\frac{y+q^2/s}{1-y-q^2/s}\rho(q^2).
\label{funfl}
\end{equation}
The distribution of the effective electron antineutrinos in the positron is also given by~\eqref{funw1} and~\eqref{funfl} due to $CP$ invariance. The presented results apply to the distributions of the other neutrinos, $\nu_{\mu}$ in $\mu^-$ and $\nu_\tau$ in $\tau^-$, provided that the mass of the given parent charged lepton is negligibly small compared to the reaction energies and the masses of the boson ($m^2_{\mu,\tau}\ll s$, $m_{\mu,\tau}^2\ll q^2$).

\section{Application to $e^+e^-$ collisions \label{sec2}}
\subsection{The Glashow resonance}
The Standard Model includes processes the existence of which is not yet convincingly proved in experiments.  This is the case for the $s$-channel resonant  production of the  on-shell $W^-$ boson in 

\begin{equation}
\bar\nu_e e^-\rightarrow W^-. \label{Glashow_r}
\end{equation}
This channel was predicted by Glashow in 1959~\cite{Glashow_res} and is now usually referred to as the Glashow resonance.

\begin{figure}[H]
\centerline{\includegraphics[width=0.3\textwidth]{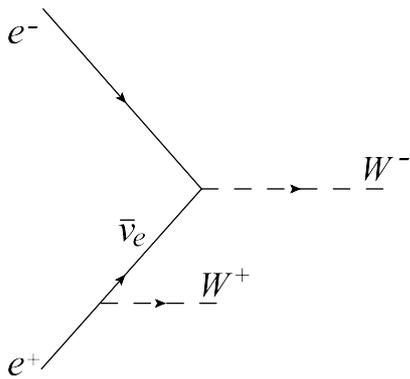}}
\caption{The Glashow resonance in the reaction $e^+e^-\rightarrow W^+W^-$.}
\label{fig2p}
\end{figure}

The strategy of observation of the resonance adopted today is to search for it in large volume water/ice neutrino detectors by exploiting the flux of very high-energy cosmic antineutrinos~\cite{berezinsky}.
The reaction $\bar\nu_e e^-\rightarrow W^-$ in the context of the effective electron approximation has been also considered in~\cite{Alikhanov:2008eu,Alikhanov:2014uja,Alikhanov:2015kla}. 
Below we show that in the framework of the ENA the Glashow resonance can be probed at electron--positron colliders.  

The resonance will be excited when the incident electron annihilates on an effective electron antineutrino from the positron as illustrated in Fig.~\ref{fig2p}. The $CP$ conjugate resonance, $\nu_ee^+\rightarrow W^+$, will appear similarly. The whole process will be observed in the following form: 

\begin{equation}
e^+ e^-\rightarrow W^+W^-. \label{r_ww}
\end{equation}
Note that only the transversely polarized bosons contribute to~\eqref{r_ww}. Then, in analogy to the QPM, the cross section for~$e^+ e^-\rightarrow W^+W^-$ will be given as a convolution: 

\begin{equation}
\sigma(s)=2\int_{0}^{\left(1-m_W/\sqrt{s}\right)^2}f^T_{\nu/e}(y,Q^2)\sigma_{\nu e\rightarrow W}(ys)\mathrm{d}y,\label{main_sec}
\end{equation}
where $\sigma_{\nu e\rightarrow W}(s)$ is the cross section for the subprocesses $\bar\nu_ee^-\rightarrow W^-$ and $\nu_ee^+\rightarrow W^+$. The factor 2 implies that the distributions for $\nu_e$ in $e^-$ and $\bar\nu_e$ in $e^+$ are equal to each other because we assume $CP$ invariance. In order to evaluate the integral of~\eqref{main_sec} in the resonance region we use the narrow width approximation: 

\begin{equation}
\sigma_{\nu e\rightarrow W}(s)=\frac{2\pi^2\alpha}{\sin^2{\theta_W}}\delta(s-m_W^2). \label{narrow}
\end{equation}
Substituting~\eqref{funw} and~\eqref{narrow} into~\eqref{main_sec} and taking into account that at $s\rightarrow\infty$ for reaction~\eqref{r_ww},  $(Q^2_{\text{max}}+m_W^2)/(Q^2_{\text{min}}+m_W^2)\rightarrow s/m_W^2$, we obtain

\begin{equation}
\sigma(s)=\frac{\pi\alpha^2}{2\sin^4{\theta_W}}\frac{1}{s}\ln\left(\frac{s}{m_W^2}\right).\label{main_secf2}
\end{equation}
Figure~\ref{fig4pp} shows the cross section as a function of the cms energy. 
It should be emphasized that~\eqref{main_secf2} coincides with

\begin{figure}[H]
\centerline{\includegraphics[width=0.3\textwidth]{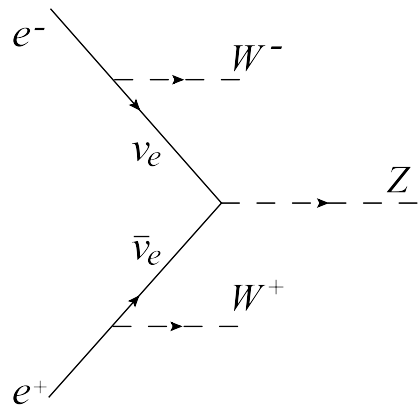}}
\caption{The $\nu_e\bar\nu_e\rightarrow Z$ annihilation in $e^+e^-$ collisions.}
\label{fig3p}
\end{figure}

\noindent the exact cross section derived within the electroweak theory in the asymptotic limit~\cite{Flambaum:1974wp,Alles:1976qv}. This is because the logarithmic part of the exact cross section comes from the terms, in the amplitude squared, behaving as $\sim1/t$ (with $t=-Q^2$ being  the Mandelstam variable). One can check that each term of this type is related to the Feynman diagram containing the neutrino~\cite{Alles:1976qv}. This is either the square of the diagram itself or an interference term with this diagram.  
Our model provides thus an interpretation of the nature of the reaction $e^+e^-\rightarrow W^+W^-$: it is asymptotically a sum of the Glashow resonance and its $CP$ conjugate.
Already at the cms energy $\sqrt{s}=500$~GeV the contribution of the non-resonant channels (with the $\gamma$ and $Z$ boson exchanges) to $e^+e^-\rightarrow W^+W^-$ drops down to the level of a few percent and continues decreasing at TeV energies. Such energies can be achieved at future electron--positron colliders as the International Linear Collider (ILC)~\cite{Fujii:2015jha} and the Compact Linear Collider (CLIC)~\cite{CLIC:2016zwp}. For instance, the integrated luminosity  of~4000~fb$^{-1}$ expected at the ILC for $\sqrt{s}=500$~GeV~\cite{Barklow:2015tja}  would allow to produce $\sim3\times10^7$ Glashow resonance events. The CLIC can study this reaction at substantially higher energies in its second and third stages of operation with the cms energies 1.5~TeV and 3~TeV, respectively. The corresponding integrated luminosities are planned to be~2.5~ab$^{-1}$ and 5~ab$^{-1}$~\cite{Robson:2018zje}, so that the CLIC will be capable of producing more than $10^{6}$ events in each of the stages. Exploiting specific observation channels along with the limitations of experimental apparatus will certainly reduce these numbers. For example, in the relatively clear channel with two uncorrelated oppositely charged muons in the final state (due to the subsequent $W^{\pm}\rightarrow\mu^\pm+\nu$ decays), one would observe approximately the total event rate times $(\Gamma_{W^+\rightarrow~\mu^+\nu}/\Gamma_{W^+\rightarrow~\text{all}})^2\approx10^{-2}$.  

\subsection{The $Z$-burst mechanism}

Another example is annihilation of neutrino--antineutrino pairs into the neutral electroweak boson:

\begin{equation}
\nu\bar\nu\rightarrow Z.
\end{equation}

\begin{figure}[t]
\centerline{\includegraphics[width=0.48\textwidth]{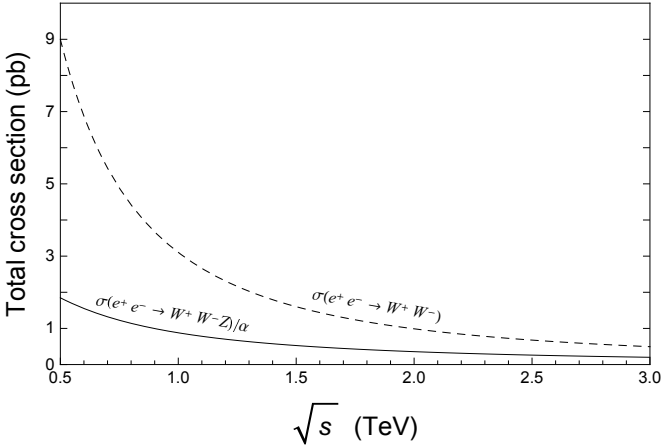}}
\caption{Total cross sections for the Glashow resonance excitation in $e^+e^-\rightarrow W^+W^-$ given by~\eqref{main_secf2} (dotted curve) and for the $Z$-burst channel in $e^+e^-\rightarrow W^+W^-Z$,~\eqref{main_secfz},  divided by $\alpha$ (solid curve).}
\label{fig4pp}
\end{figure}

\noindent This process has an important implication for cosmology because provides a tool to probe the relic neutrino background~\cite{Weiler:1982qy} but still remains unidentified experimentally. Ultra-high-energy neutrinos scattering onto relic neutrinos may excite the $Z$ resonance which, subsequently decaying, would  produce a shower of observable secondary particles. It is sometimes called the $Z$-burst mechanism. The $\nu_e\bar\nu_e\rightarrow Z$ channel of this scenario can be tested at $e^+e^-$ colliders in the following form (see Fig.~\ref{fig3p}):

\begin{equation}
e^+e^-\rightarrow W^+W^-+Z. \label{burst_mf}
\end{equation}
Within the ENA the cross section for~\eqref{burst_mf} is represented as

\begin{equation}
\sigma_{Z}(s)=\int \int f^T_{\nu/e}(x,Q^2)f^T_{\nu/e}(y,Q^2)\sigma_{\nu \bar\nu\rightarrow Z}(xys)\mathrm{d}x\mathrm{d}y,\label{main_secz}
\end{equation}
where $\sigma_{\nu \bar\nu\rightarrow Z}(s)$ is the cross section for the subprocess $\nu_e\bar\nu_e\rightarrow Z$. 
The narrow width approximation gives $\sigma_{\nu \bar\nu\rightarrow Z}(s)=8\pi^2\alpha\,\delta(s-m_Z^2)/\sin^2{2\theta_W}$, so that ~\eqref{main_secz} asymptotically behaves as

\begin{equation}
\sigma_Z(s)\propto\frac{\alpha^3\cos^4{\theta_W}}{\sin^6{2\theta_W}}\frac{1}{s}\ln^2\left(\frac{s}{m_W^2}\right).\label{main_secfz}
\end{equation}
This cross section as a function of the cms energy is plotted in Fig.~\ref{fig4pp}. The CLIC in the third energy stage at $\sqrt{s}=3$~TeV  is capable of producing more than $10^4$ $Z$-burst events.

\begin{figure}[H]
\centering
\centerline{\includegraphics[width=0.35\textwidth]{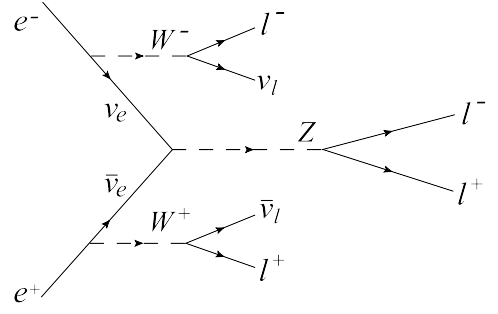}}
\caption{The 4-lepton signal ($l^+l^-+l^+l^-$) for the $\nu_e\bar\nu_e\rightarrow Z$ annihilation in $e^+e^-$ collisions.}
\label{fig4p}
\end{figure}

Due to the leptonic decay modes of the final bosons, a useful signature of the appearance of an off-shell $Z$ in the $\nu_e\bar\nu_e$ channel is two charged lepton pairs in the final state:

 \begin{equation}
 e^+e^-\rightarrow W^+W^-+Z\rightarrow {\it l}^+{\it l}^-+{\it l}^+{\it l}^-  
\end{equation}
with one of the lepton pairs being highly massive (see Fig.~\ref{fig4p}). This is similar to the observation of massive $\mu^+\mu^-$ pairs in $pp$ collisions through Drell--Yan processes. Note 

\begin{figure*}
\centerline{\includegraphics[width=0.8\textwidth]{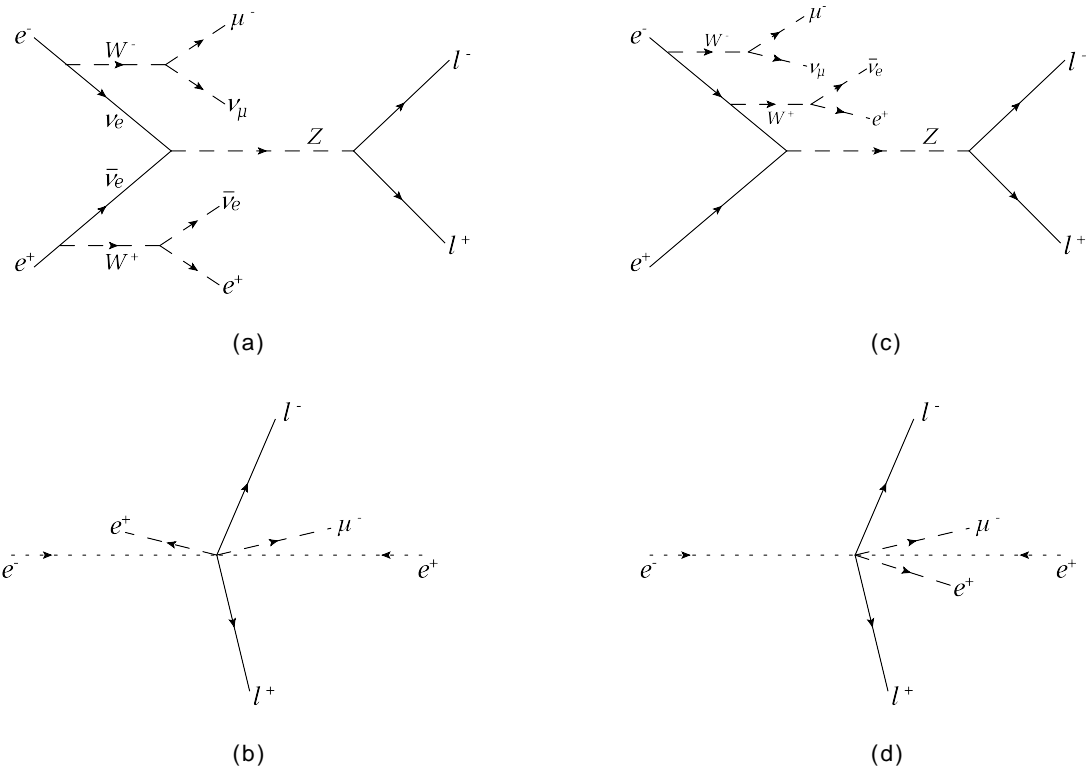}}
\caption{An example of a 4-lepton signal for the $\nu_e\bar\nu_e\rightarrow Z$ annihilation in $e^+e^-$ collisions (a) and its signature in a detector (b). The diagram of the corresponding background event and its signature in the  detector are shown by (c) and (d), respectively. In (b) and (d), azimuthal symmetry around the beam axis, denoted by the dotted line, is assumed.}
\label{fig5p}
\end{figure*}

\begin{figure}[H]
\centerline{\includegraphics[width=0.35\textwidth]{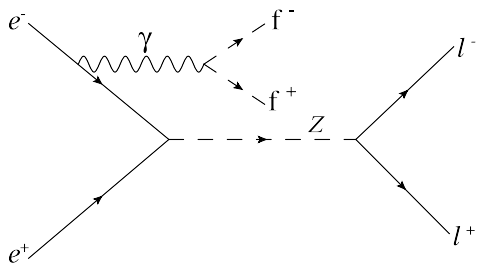}}
\caption{An example of a diagram for a process with photon conversion into a fermion--antifermion pair that may also contribute to the background for the $\nu_e\bar\nu_e\rightarrow Z$ annihilation events. Its signature in a detector will be similar to that in Fig.~\ref{fig5p}(d).}
\label{fig6p}
\end{figure}

\noindent that the leptons from the $W^\pm$ boson decays are not necessarily from the same generation, so that a distinctive signal may be, for example, of the form $e^{\pm}\mu^{\mp}+l^+l^-$, as depicted in Figs.~\ref{fig5p}(a) and~(b). Thus the signal in a detector will be visible as an uncorrelated electron and a muon with longitudinal momenta of opposite signs (from the $W^{\pm}\rightarrow l^{\pm}+\nu$ decays) and two leptons of the same generation with balanced transverse momenta (from the $Z\rightarrow l^+l^-$ decays). In contrast, the background events, shown in Figs.~\ref{fig5p}(c) and~(d), will have a different signature and may be therefore reduced. This is because in a background event, at $\sqrt{s}\gg m_W$, both $W^+$ and $W^-$ will be emitted either by an electron or by a positron predominantly along the beam axis  and decaying will give the charged leptons with longitudinal momenta of the same sign. The background from the photon conversion into charged fermion pairs, Fig.~\ref{fig6p}, will have the same behavior. The latter must be taken into account provided the given experiment cannot distinguish between leptons of different generations. The quark decay modes of the bosons can also be used to identify the $Z$-burst events. In this case the signal will contain two jets with balanced transverse momenta plus two jets with uncorrelated momenta. The states with jets and charged leptons may also be combined. The background can be isolated again due to the peculiar distribution of the longitudinal components of the jet momenta. The Feynman-$x$ variable, $x_F=2p^*_{||}/\sqrt{s}$, should be useful in this regard ($p^*_{||}$ is the longitudinal component of the particle momentum in the cms frame).

It is notable that models with the so-called "secret neutrino interactions" with hypothetical new mediators coupling to neutrinos~\cite{Kolb:1987qy} can be similarly tested at high momentum transfers in electron--positron collisions.

\section{Conclusions \label{concl}}
Neutrinos are likely the most poorly understood basic constituents of the Standard Model. It is a challenging problem to create high intensity and well collimated neutrino beams with known flavor compositions for related experiments.
In this article we have proposed the effective neutrino method as a framework for studying neutrino interactions. A charged lepton is able to manifest 

\noindent itself, with a certain probability, as a neutrino. We have derived the distributions of the effective neutrinos in the charged leptons. This is analogous to the parton densities in hadrons introduced in order to investigate dynamical properties of quarks and gluons unavailable in free states. 
The method  may provide an additional tool for probing neutrino-induced reactions at $e^+e^-$ and $ep$ colliders as well as at other facilities that use charged lepton beams. For example, we have shown that i) the reaction $e^+e^-\rightarrow W^+W^-$ asymptotically is determined by the Glashow resonance and its $CP$ conjugate; and ii) the $\nu\bar\nu\rightarrow Z$ annihilation can be observed in the channel $e^+e^-\rightarrow W^+W^-Z$. Both the Glashow resonance and the $Z$-burst mechanism remain so far unidentified in traditional experiments exploiting ultra-high-energy neutrino component of cosmic rays, however can be accessed, according to our study, at electron--positron colliders operating at reasonable center-of-mass energies.

The proposed approach is simple and can be readily applied to other neutrino types as well as to neutrino--quark interactions. For example, high massive final states in $\nu_{e}q$ and $\nu_{\mu}q$ scattering processes in $ep$ and $\mu p$ collisions can be treated in similar way. 
In closing, we point out that the effective neutrino method has at least the following advantages:
\vspace{-0.2cm}

\begin{itemize}[leftmargin=0.75cm,labelsep=0.15cm]
 \setlength\itemsep{0.13cm}
\item[1)] Various neutrino-induced processes, observation of 
which in the laboratory conditions 
is considered by many 
unrealistic or at least very difficult, become 
accessible with ordinary collider experiments.
\item[2)] Techniques of charged particle beam acceleration are
well established.
\item[3)] The collision energy can be varied in a wide range.
\item[4)] High luminosities can be achieved.
\item[5)] The interaction region is localized.
\end{itemize}

\acknowledgments
I would like to thank Professor E. A. \Name{Paschos} for reading the manuscript and encouraging comments. This work was partly supported by the Program of fundamental scientific research of the Presidium of the Russian Academy of Sciences "Physics of fundamental interactions and nuclear technologies".




\begin{thebibliography}{99}

\bibitem{King:2014nza} 
  \Name{King S. F., Merle A., Morisi S., Shimizu Y. \and M.~Tanimoto}
  \REVIEW{New J. Phys.}{16}{2014}{045018}.
	
	
\bibitem{Bilenky:2014ema} 
  \Name{Bilenky S. M.}
 \REVIEW{Phys. Part. Nucl.}{46}{2015}{475}.


\bibitem{Fukuda:1998ah} 
  \Name{Fukuda Y. {\it et al.} [Super-Kamiokande Collaboration]}
  \REVIEW{Phys. Rev. Lett.}{82}{1999}{2644}.
	
\bibitem{Ahmad:2002jz} 
  \Name{Ahmad Q. R. {\it et al.} [SNO Collaboration]}
  \REVIEW{Phys. Rev. Lett.}{89}{2001}{011301}.
	
\bibitem{Eguchi:2002dm} 
  \Name{Eguchi K. {\it et al.} [KamLAND Collaboration]}
  \REVIEW{Phys. Rev. Lett.}{90}{2003}{021802}.
	
\bibitem{Zucchelli:2002sa} 
  \Name{Zucchelli P.}
 \REVIEW{Phys. Lett. B}{532}{2002}{166}.
	
\bibitem{Volpe:2003fi} 
  \Name{Volpe C.}
  \REVIEW{J. Phys. G}{30}{2004}{L1}.
	
\bibitem{Bjorken:1969ja} 
  \Name{Bjorken J. D. \and Paschos E. A.}
  \REVIEW{Phys. Rev.}{185}{1969}{1975}.


\bibitem{Brodsky:1971ud} 
  \Name{Brodsky S. J., Kinoshita T. \and Terazawa H.}
  \REVIEW{Phys. Rev. D}{4}{1971}{1532}.

\bibitem{Walsh:1973mz} 
  \Name{Walsh T. F. \and Zerwas P. M.}
  \REVIEW{Phys. Lett. B}{44}{1973}{195}.



\bibitem{Budnev:1974de} 
  \Name{Budnev V. M., Ginzburg I. F., Meledin G. V. \and Serbo V. G.}
  \REVIEW{Phys. Rept.}{15}{1975}{181}.
	
\newpage		
	
\bibitem{Kane:1984bb} 
  \Name{Kane G. L., Repko W. W. \and Rolnick W. B.}
  \REVIEW{Phys. Lett. B}{148}{1984}{367}.


\bibitem{Dawson:1984gx} 
  \Name{Dawson S.}
  \REVIEW{Nucl. Phys. B}{249}{1985}{42}.


	
\bibitem{Drell:1970wh} 
  \Name{Drell S. D. \and Yan T. M.}
  \REVIEW{Phys. Rev. Lett.}{25}{1970}316.
	
\bibitem{Frixione:1993yw} 
  \Name{Frixione S., Mangano M. L., Nason P. \and Ridolfi G.}
  \REVIEW{Phys. Lett. B}{319}{1993}{339}.
	
	
\bibitem{Chen:1975sh} 
  \Name{Chen M. S. \and Zerwas P. M.}
  \REVIEW{Phys. Rev. D}{12}{1975}{187}.
	
\bibitem{Alikhanov:2018kpa} 
  \Name{Alikhanov I.}
  arXiv:1812.05578 preprint, 2018.
	
	
	\bibitem{Glashow_res}
\Name{Glashow S. L.}
  \REVIEW{Phys. Rev.}{118}{1960}316.
	
	\bibitem{berezinsky}
  \Name{Berezinsky V. S. \and Gazizov A. Z.}
  \REVIEW{JETP Lett.}{25}{1977}254.

\bibitem{Alikhanov:2008eu} 
  \Name{Alikhanov I.}
  \REVIEW{Eur. Phys. J. C}{56}{2008}479.
	
 \bibitem{Alikhanov:2014uja} 
 \Name{Alikhanov I.}
  \REVIEW{Phys. Lett. B}{741}{2015}{295}.


\bibitem{Alikhanov:2015kla} 
 \Name{Alikhanov I.}
	\REVIEW{Phys. Lett. B}{756}{2016}{247}.
	
  \bibitem{Flambaum:1974wp} 
  \Name{Flambaum V. V., Khriplovich I. B. \and Sushkov O. P.}
  \REVIEW{Sov. J. Nucl. Phys.}{20}{1975}{537}.
	
\bibitem{Alles:1976qv} 
  \Name{Alles W., Boyer C. \and Buras A. J.}
  \REVIEW{Nucl. Phys. B}{119}{1977}{125}.

\bibitem{Fujii:2015jha} 
  \Name{Fujii K. {\it et al.}} ILC-NOTE-2015-067,
  arXiv:1506.05992 preprint, 2015.


\bibitem{CLIC:2016zwp} 
\Name{Burrows P. N. {\it et al.} [CLICdp and CLIC Collaborations]}
  \REVIEW{CERN Yellow Rep. Monogr.}{1802}{2018}{1}.

\bibitem{Barklow:2015tja} 
 \Name{Barklow T., Brau J., Fujii K., Gao J., List J., Walker N. \and Yokoya K.}
  arXiv:1506.07830 preprint, 2015.

\bibitem{Robson:2018zje} 
  \Name{Robson A. \and Roloff P.} CLICdp-Note-2018-002,
  arXiv:1812.01644 preprint, 2018.
  


\bibitem{Weiler:1982qy} 
  \Name{Weiler T. J.}
  \REVIEW{Phys. Rev. Lett.}{49}{1982}{234}.

\bibitem{Kolb:1987qy} 
  \Name{Kolb E. W. \and Turner M. S.}
  \REVIEW{Phys. Rev. D}{36}{1987}{2895}.
	

	\end{thebibliography}
\end{document}